\documentclass[epj,twocolumn]{webofc}
\woctitle{ISVHECRI 2018}
\usepackage[varg]{txfonts}
\usepackage{lineno}

\begin{document}

\title{Future ground arrays for ultrahigh-energy cosmic rays: recent updates and perspectives}

\author{\firstname{Toshihiro} \lastname{Fujii}\inst{1}\fnsep\thanks{\email{fujii@icrr.u-tokyo.ac.jp}}}

\institute{Institute for Cosmic Ray Research, University of Tokyo, Kashiwa, Chiba, Japan}

\abstract{
The origin and nature of ultrahigh-energy cosmic rays (UHECRs) are one of the most intriguing mysteries in particle astrophysics and astronomy. 
The two largest observatories, the Pierre Auger Observatory and the Telescope Array Experiment, are steadily observing UHECRs in both hemispheres 
in order to better understand their origin and associated acceleration mechanisms at the highest energies.
We highlight their latest results including on-going upgrades, AugerPrime and TA$\times$4, and then address the requirements for a next-generation observatory.
We share recent updates and perspectives for a future ground array of fluorescence detectors, addressing the requirements for a large-area, 
low-cost detector suitable for measuring the properties of the highest energy cosmic rays with an unprecedented aperture.
}

\maketitle

\section{Ultrahigh-energy cosmic rays}
\label{intro}
In 1912, V.\,F. Hess discovered a flux of energetic elementary particles arriving from outer space, now known as ``Cosmic Rays'' \cite{Hess:1912srp}.
The cosmic-ray energy spectrum has been measured at energies from 10$^{8}$\,eV to beyond 10$^{20}$\,eV in the last century, and follows a power-law relationship of the form $E^{-3}$, where $E$ is the cosmic ray energy.
Cosmic rays above 10$^{18}$\,eV (= 1 EeV) are known as ultrahigh-energy cosmic rays (UHECRs).

Since UHECRs are the most energetic particles in the universe, their origins are ostensibly related to extremely energetic astrophysical phenomena, such as gamma-ray bursts, active galactic nuclei, or other exotic processes such as the decay or annihilation of super-heavy relic particles created in an early phase of the development of the universe \cite{Hillas:1985is}.
However, their origin and acceleration mechanism above 10$^{20}$\,eV are still unknown. Therefore, they are one of the most intriguing mysteries in particle astrophysics and astronomy. 

\subsection{GZK cutoff and charged particle astronomy in nearby universe}
Following the detection of a cosmic-ray event with an energy of 10$^{20}$\,eV by J. Linsley \cite{Linsley:1963km} in 1963, 
K. Greisen, G.\,T. Zatsepin and V.\,A. Kuzmin predicted the UHECR energy spectrum to be suppressed above 10$^{19.7}$\,eV due to the interaction of high-energy particles with the 3\,K cosmic microwave background radiation via pion production, the so-called GZK cutoff \cite{bib:gzk1, bib:gzk2}. 
If the GZK cutoff exists, the origin of UHECRs is significantly restricted to nearby sources distributed non-uniformly within 50-100\,Mpc. 
Additionally, UHECRs propagate with less deflection by magnetic fields due to their enormous kinetic energies. 
As a result, the arrival directions of UHECRs should be correlated with the directions of extremely energetic sources or objects, leading to the possibility of next-generation particle astronomy with UHECRs, illuminating extremely energetic phenomena in the nearby universe.

\subsection{Detection methods and UHECR observatories}
\label{method}
\begin{figure}[tbh]
\centering
\includegraphics[width=0.8\linewidth,clip]{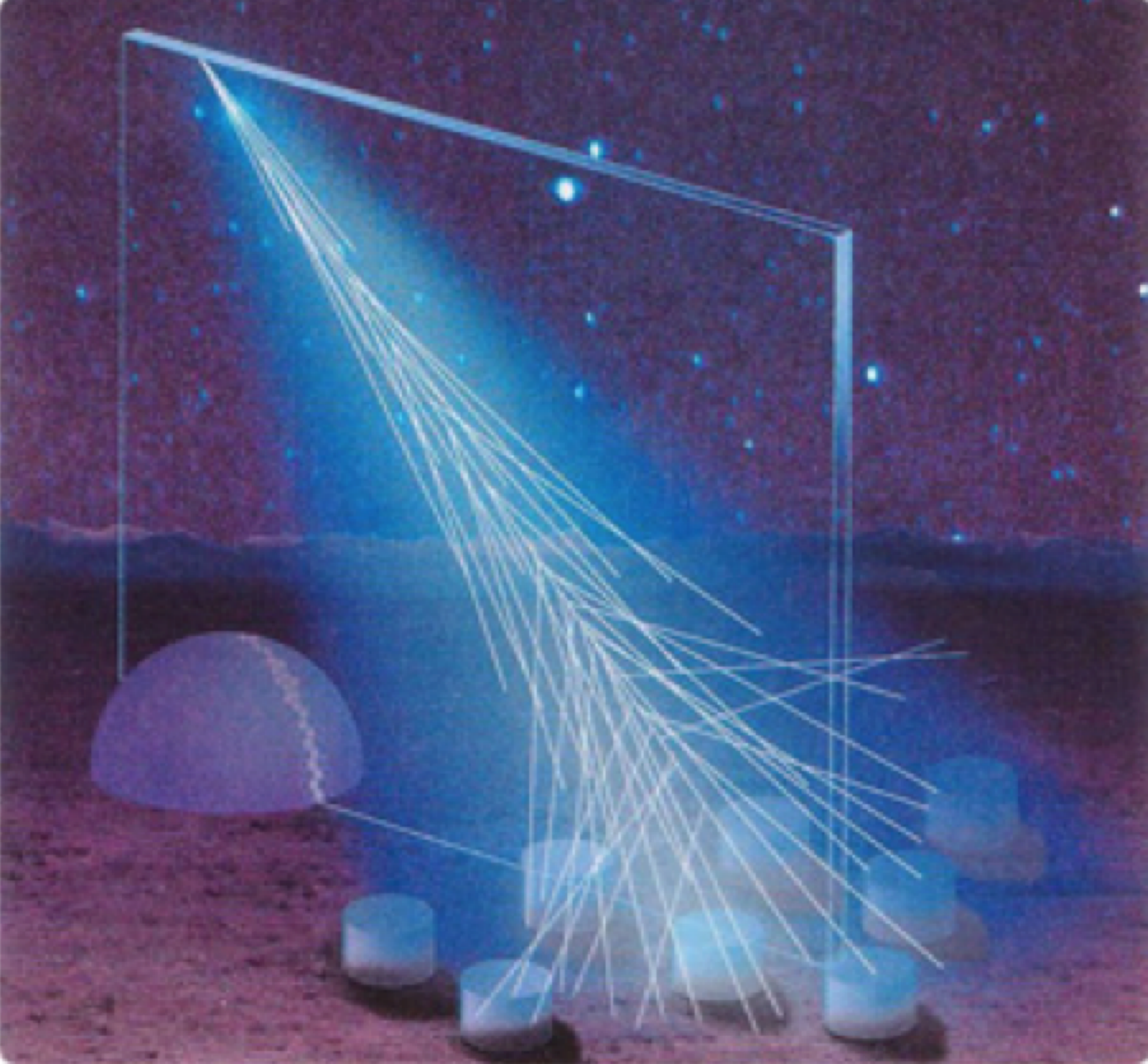}
\caption{Schematic view of  arrays of detectors and large-field-of-view fluorescence telescopes, detecting signals from EAS induced by UHECRs [https://www.auger.org].}
\label{fig:method}
\end{figure}
\begin{figure*}[tbh]
\centering
\includegraphics[width=1.0\linewidth,clip]{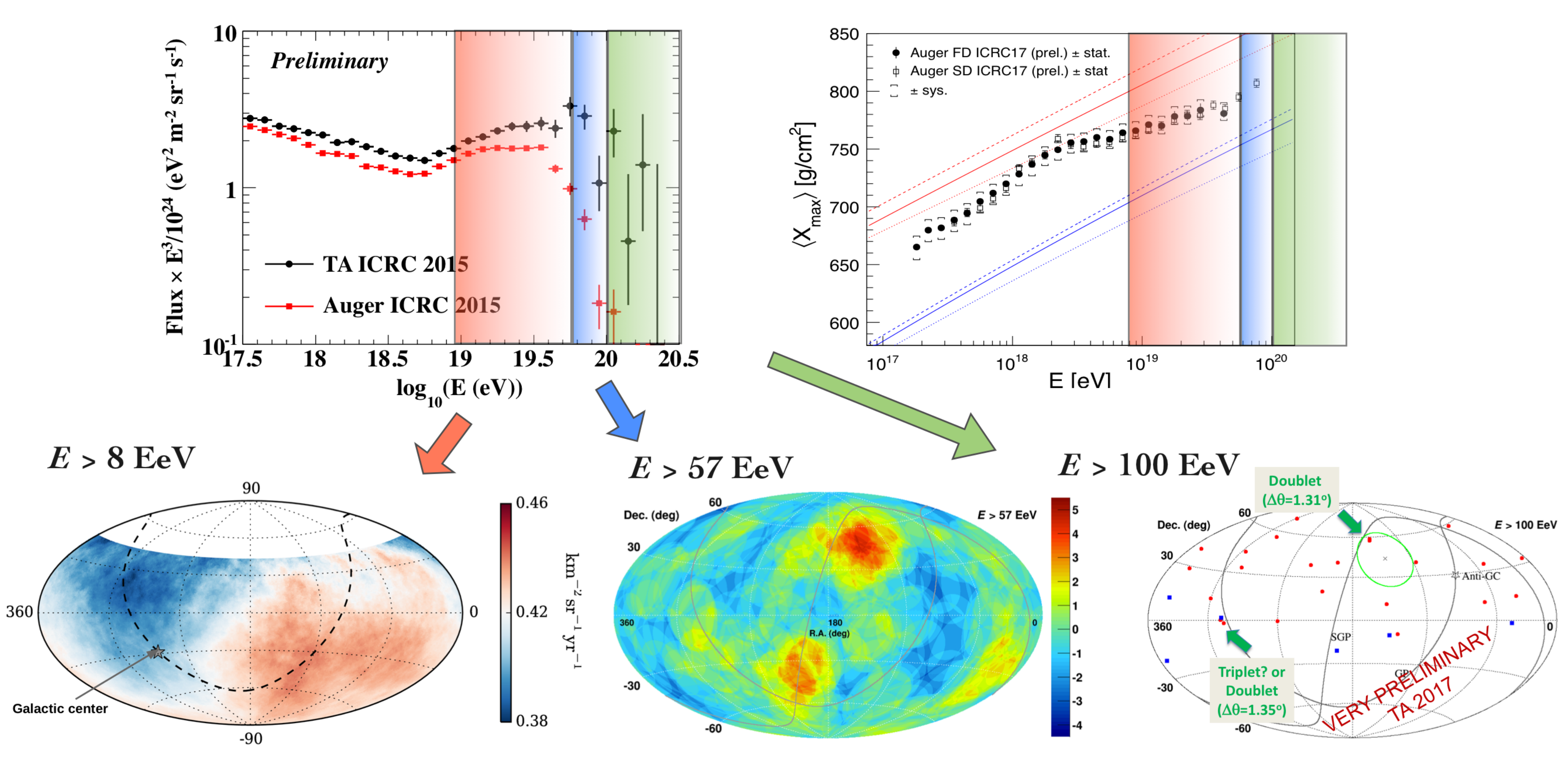}
\caption{Overview of three important observables reported by the Pierre Auger Observatory and Telescope Array Experiment: energy spectrum \cite{Valino:2015zdi, bib:taspectrum_icrc2015, Verzi:2017hro}  mass composition \cite{bib:mass_auger,bib:mass_implication_auger,bib:delta_auger} and arrival direction \cite{bib:dipole_auger, bib:hotspot_ta, Kawata:2015whq, Troitsky:2017xqh} with energy ranges indicated above: 8 EeV (red), 57 EeV (blue) and 100 EeV (green).}
\label{fig:overview}
\end{figure*}

Given the minute flux of UHECRs, less than one particle per century per square kilometer at the highest energies, a very large area must be instrumented in order to collect significant statistics. 
The energy, arrival direction, and mass composition of UHECRs can be inferred from studies of the cascades of secondary particles (Extensive Air Shower, EAS) produced by their interaction with the Earth's atmosphere. 

Two well-established techniques are used for UHECR detection: arrays of detectors (e.g. plastic scintillators, water-Cherenkov stations) sample EAS particles reaching the ground; large-field-of-view telescopes allow for reconstruction of the shower development in the atmosphere by imaging ultraviolet fluorescence light from atmospheric nitrogen excited by EAS particles as shown in Figure \ref{fig:method}.
Two giant observatories, one in each hemisphere, the Pierre Auger Observatory (Auger) in Mendoza, Argentina \cite{bib:auger} and the Telescope Array Experiment (TA) in Utah, USA \cite{bib:tafd, bib:tasd}, combine the two techniques with arrays of particle detectors overlooked by fluorescence detectors (FD).

\begin{figure*}[tbh]
\centering
\includegraphics[width=1.0\linewidth,clip]{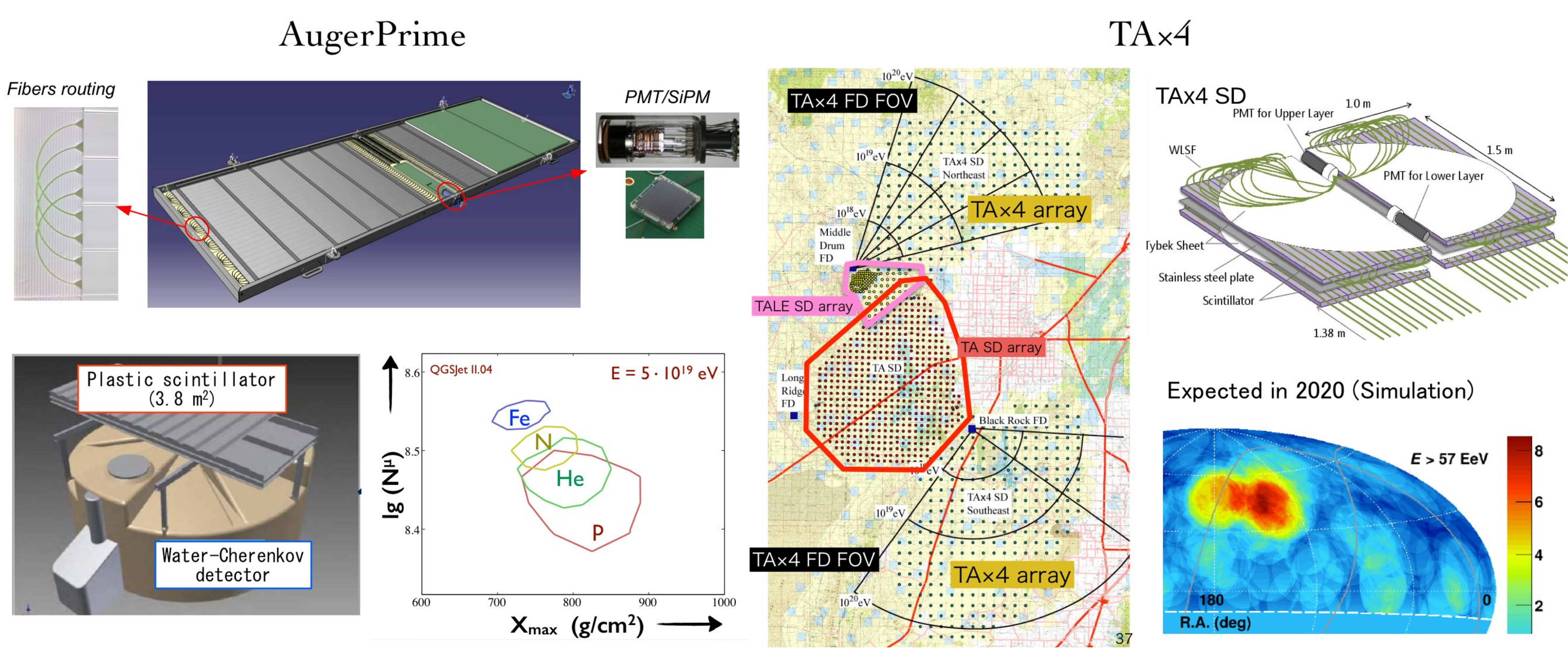}
\caption{Ongoing upgrades of AugerPrime (left) \cite{bib:augerprime} and TA$\times$4 (right)\cite{bib:tax4}. AugerPrime will provide excellent sensitivity to the mass composition using a measurement of the number of muons $N_{\mu}$ and the depth of shower maximum $X_{\max}$ from the surface detector array. TA$\times$4 will provide a four-fold increase in statistics compared to TA, allowing for a detailed analysis of the TA hot-spot. }
\label{fig:upgrades}
\end{figure*}

\section{Latest results from current UHECR observatories}
\label{results}
Figure \ref{fig:overview} shows the latest results on the energy spectrum, arrival directions and mass composition of UHECRs reported by the Pierre Auger Observatory and the Telescope Array Experiment~\cite{Valino:2015zdi,bib:taspectrum_icrc2015,Verzi:2017hro,bib:mass_auger,bib:mass_implication_auger,bib:delta_auger,bib:dipole_auger,bib:hotspot_ta,Kawata:2015whq,Troitsky:2017xqh}.
Both observatories have measured a dip around 10$^{18.7}$\,eV and a suppression above 10$^{19.7}$\,eV~\cite{Valino:2015zdi,bib:taspectrum_icrc2015}.
The suppression is consistent with expectation from the GZK cutoff, however the shape of the suppression is discrepant between the two measurements~\cite{Verzi:2017hro}. 

The mass composition reported by Auger through $X_{\max}$ (the depth in the atmosphere at which the EAS reaches its maximum energy deposit) suggests a transition from light nuclei at around 10$^{18.3}$\,eV to heavier nuclei up to energies of 10$^{19.6}$\,eV~\cite{bib:mass_auger, bib:mass_implication_auger}.
In addition, the $X_{\max}$ distributions measured by Auger and TA below 10$^{19}$\,eV are consistent within systematic uncertainties~\cite{bib:composition_wg, deSouza:2017wgx}.
Recently Auger reported a new mass composition result using the surface detector array, sensitive to primary energies above 57\,EeV, using measurements of the signal rise time~\cite{bib:delta_auger}. 
However, there is currently no mass composition result above 100\,EeV. 

Figure \ref{fig:overview} also includes the measured arrival directions of UHECRs above 8\,EeV, 57\,EeV and 100\,EeV, respectively.
The left sky-map is a flux map of UHECRs above 8\,EeV reported by Auger, indicating an obvious dipole structure of 6.5\% amplitude with a 5.2$\sigma$ significance~\cite{bib:dipole_auger}. The direction of maximum amplitude is 125$^{\circ}$ away from our galactic center, indicating an extragalactic origin for these ultrahigh-energy particles.
This dipole structure and its amplitude are consistent with expectation if we assume the deflection of charged nuclei under the typical strength of the magnetic fields within our galaxy~\cite{diMatteo:2017dtg}.

The central sky-map is a significance map of UHECR arrival directions above 57\,EeV observed with both TA and Auger~\cite{Kawata:2015whq}.
The maximum excess appears as a ``hotspot'' centered at a right ascension of 147$^{\circ}$ and a declination of 43$^{\circ}$ with a 3.4$\sigma$ significance~\cite{bib:hotspot_ta}. In addition, arrival directions of UHECRs are found to correlate with nearby extragalactic objects at a modest 2 $\sim$ 3$\sigma$ significance level~\cite{bib:anisotropy_ta, bib:anisotropy_auger}. 
There seems to be no measured excess in the direction of the Virgo cluster.
Recently, Auger reported a 4.0$\sigma$ correlation between the positions of nearby starburst galaxies and the arrival directions of 9.7\% of their measured UHECR events above 39\,EeV~\cite{bib:sbg_auger}.
The remaining 90.3\% of their measured events are consistent with an isotropic distribution.
The right sky-map shows the arrival directions of UHECRs above 100\,EeV. Curiously, two doublets (events having arrival directions within their corresponding uncertainties) can be seen, having a chance probability of 2.8$\sigma$~\cite{Troitsky:2017xqh}. 

Except for the Auger dipole result above 8\,EeV, these inconclusive results above 57\,EeV are limited by statistics at the highest energies due to the flux suppression. 
To further advance and establish the field of charged particle astronomy, a future ground array will require an unprecedented aperture, 
which is larger by an order of magnitude, and mass composition sensitivity above 100\,EeV.

\section{Ongoing upgrades: AugerPrime and TA$\times$4}
\label{intro}
AugerPrime, shown in Figure~\ref{fig:upgrades}, is an upgrade currently underway to increase Auger's sensitivity to the mass composition of UHECRs by utilising information from plastic scintillators installed atop the surface detectors~\cite{bib:augerprime, Aab:2016vlz}.
The new 3.8\,m$^2$ scintillators will be installed to separate the electromagnetic and muonic components of measured air showers, providing increased sensitivity to the primary mass composition using the number of muons $N_{\mu}$ and the depth of shower maximum $X_{\max}$ from the surface detector array. With a factor of 10 increase in statistics compared to $X_{\max}$ measurements by fluorescence telescopes, it will be possible to search for arrival direction anisotropies with mass composition dependences. 

AugerPrime includes an update to the surface detector electronics with 120\,MHz sampling, implementation of an additional small PMT for wide dynamic range, and an extension of the duty cycle of the fluorescence detector to operate during hours of moonlight.
The buried scintillator array is also used in a ground coverage of 23.5\,km$^{2}$ to directly measure the muonic component of air showers~\cite{bib:amiga}.
The installation and data-taking of AugerPrime are scheduled to begin in 2019.

The TA$\times$4 in Figure~\ref{fig:upgrades} is a four-fold expansion in TA's ground coverage, and is aimed at collecting more events at the highest energies~\cite{bib:tax4}.
Both a surface detector array and additional fluorescence detectors will be installed. 
It will allow for the detailed mapping of the structure of the TA hotspot, with corresponding $X_{\max}$ obtained by the fluorescence detectors.
The 250 scintillators have already been assembled and shipped to Utah. Their deployment is expected in 2019.

In the next decade, Auger and TA$\times$4 will establish full-sky coverage with 3000\,km$^2$ in both the northern and southern hemispheres. 
TA$\times$4 could see the first discovery of a UHECR source, and AugerPrime could encounter a point source by selecting only events with a light composition.
After the first discovery of a UHECR source, a next-generation ground array with an effective coverage of 30,000\,km$^{2}$ will be required, with sensitivity to mass composition.

\begin{figure*}[tbh]
\centering
\includegraphics[width=1.0\linewidth,clip]{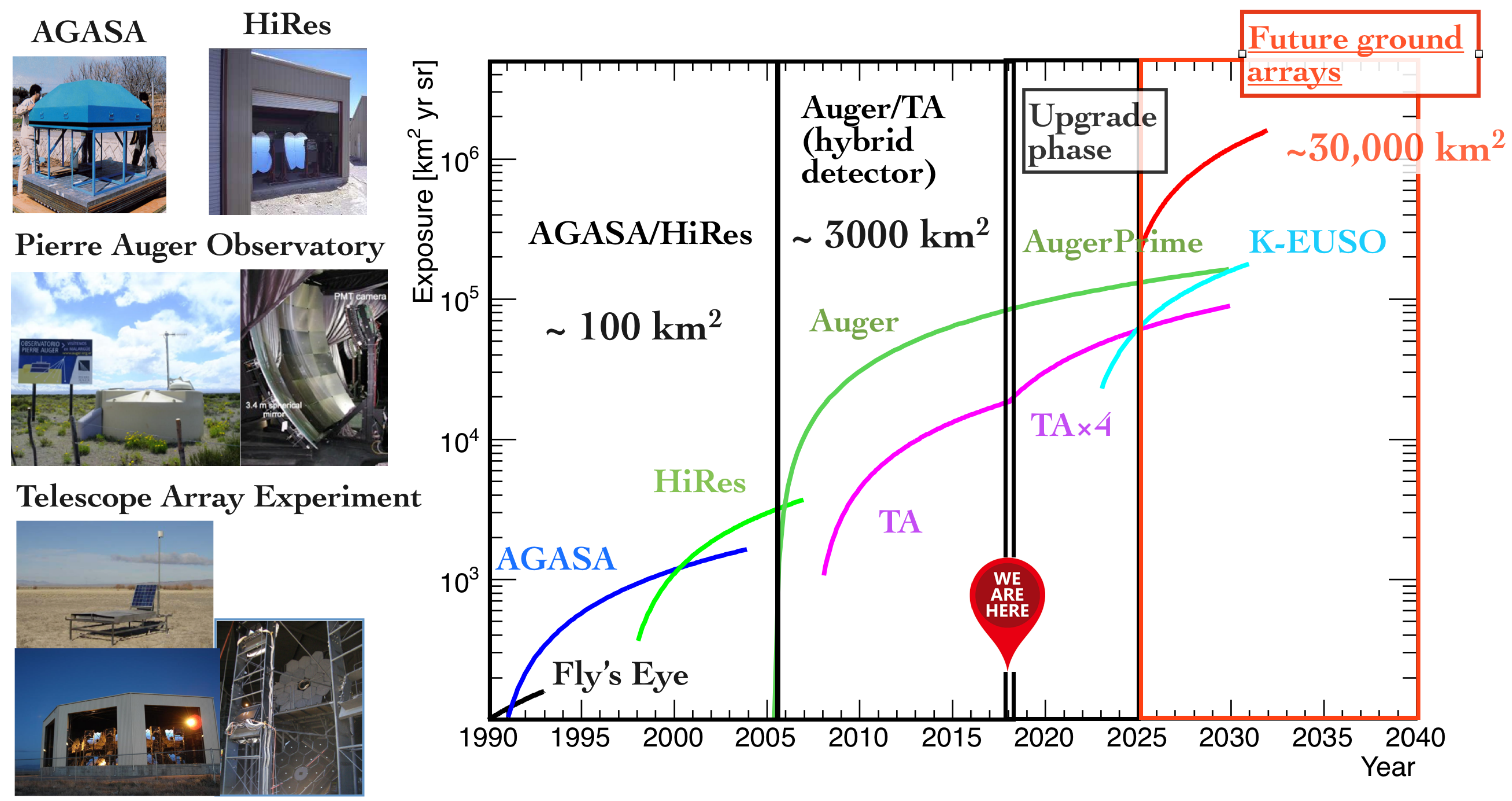}
\caption{The UHECR observatories and their exposures at the highest energies as a function of time; including AGASA \cite{bib:agasa}, High Resolution Fly's Eye \cite{bib:hires}, Auger \cite{bib:auger}, TA \cite{bib:tafd, bib:tasd}. The pioneering detection of UHECRs from space, K-EUSO \cite{bib:k-euso}, is also indicated.}
\label{fig:requirements}
\end{figure*}

\section{Requirements for a future ground array}
Figure~\ref{fig:requirements} shows the historical exposure of cosmic-ray observatories at the highest energies as a function of time.
In the 1990s, AGASA \cite{bib:agasa} and HiRes \cite{bib:hires_gzk} pioneered the detection of UHECRs using a surface detector array, and a fluorescence detector with an effective area of $\sim$100\,km$^{2}$.
In the 2000s, Auger and TA combined the two techniques, using both types of detectors simultaneously to make complementary measurements of UHECR air showers.
From now, AugerPrime and TA$\times$4 will start with a ground coverage of 3000\,km$^2$ in both hemispheres, and AugerPrime will discriminate the electromagnetic and muonic shower components for precise measurements of EAS.
Therefore, a future ground array in the 2020s requires a significant increase in the effective area, equivalent to 30,000\,km $^{2}$, and a sufficient sensitivity to the primary mass composition.

A future ground array would have the potential to observe a recovery of the GZK cutoff due to an extension of the energy spectrum beyond the suppression. 
Charged-particle astronomy will be achieved through an order of magnitude increase in statistics above 57\,EeV. Composition-dependent anisotropies would be studied using $X_{\max}$.
Also a huge target volume is needed to achieve the first detection of neutral particles such as neutrinos trionos and $\gamma$-rays at the highest energies. 
To cover the enormous target volume, future detectors should be low-cost and easily-deployable.
Worldwide collaboration is essential in successfully building such a giant ground array.

\begin{figure*}[tbh]
\centering
\includegraphics[width=1.0\linewidth,clip]{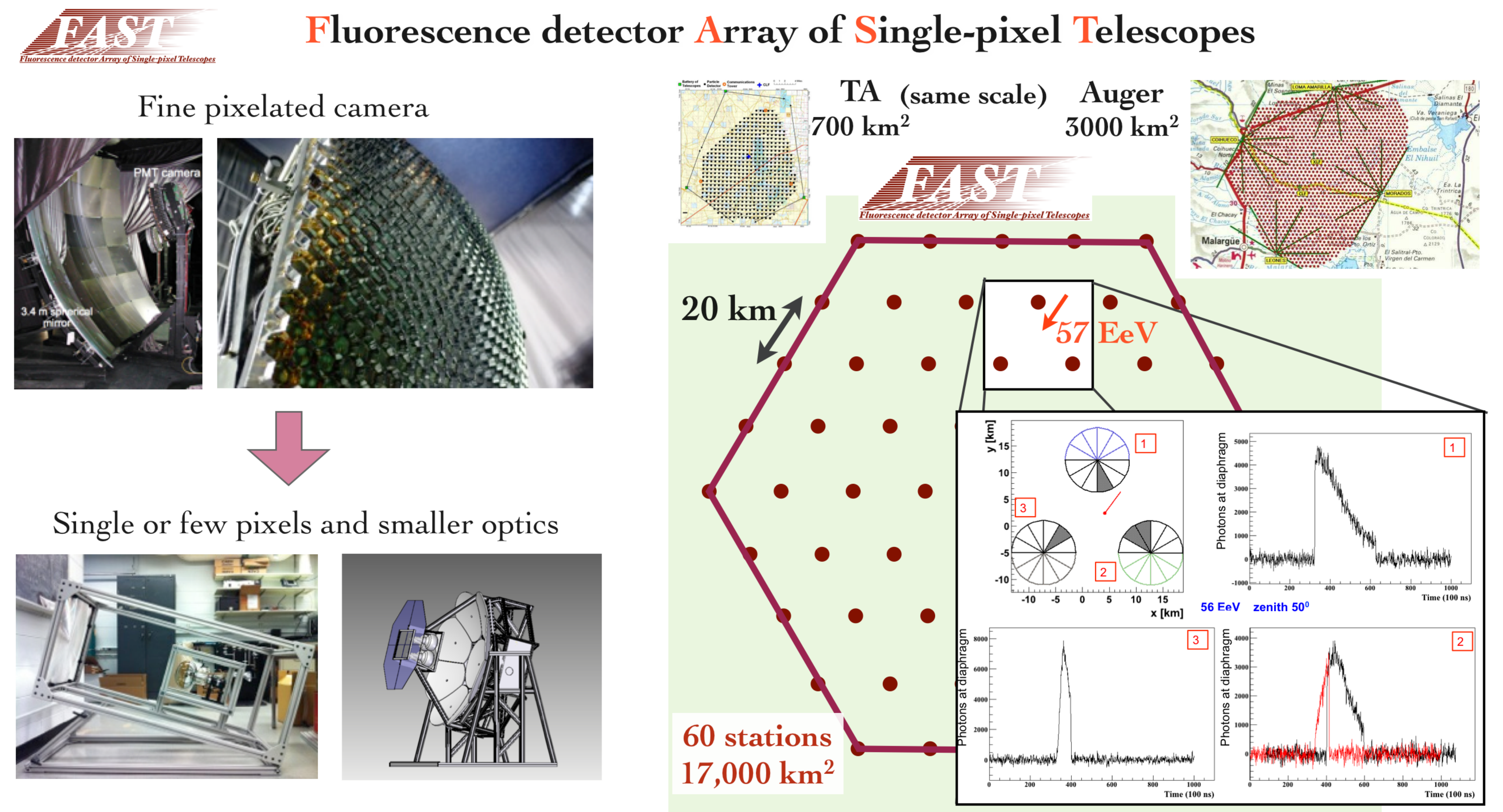}
\caption{Schematic view of the Fluorescence detector Array of Single-pixel Telescopes: one of the possible solutions for a future giant ground array~\cite{bib:fast}. As a reference, the ground coverages of TA and Auger make up only 10\% of the ground coverage of a full-scale array of FAST detectors. The expected signal emitted from a UHECR with an energy of 57\,EeV is shown in a simulation. }
\label{bib:fast}
\end{figure*}

\subsection{Fluorescence detector Array of Single-pixel Telescopes} 
One of the possible solutions capable of fulfilling these requirements is a ground-based fluorescence detector array which is low-cost, easily-deployable, and has an unprecedented aperture which is an order of magnitude larger than that of Auger and TA$\times$4. 

The Fluorescence detector Array of Single-pixel Telescopes (FAST) [https://www.fast-project.org] consists of compact FD telescopes featuring a smaller light collecting area and many fewer pixels than current generation FD designs, leading to a significant reduction in cost.

In the current FD design, a large mirror system of $\sim$3.5\,m diameter reflects a 30$^{\circ}$ $\times$ 30$^{\circ}$ patch of the sky onto a focal plane composed of several hundred photomultiplier tubes (PMTs). 
In the FAST design, the same field-of-view is covered by just four PMTs at the focal plane of a compact segmented mirror of 1.6\,m in diameter~\cite{bib:fast_optics}. 
A significant cost reduction is expected due to the compact design of FAST, with smaller light collecting optics, a smaller telescope housing, and a smaller number of PMTs and associated electronics. 
Each FAST station would consist of 12 telescopes, covering 360$^{\circ}$ in azimuth and 30$^{\circ}$ in elevation. 
An alternative optical design using Fresnel lenses was studied with the prototype optics borrowed from the EUSO-TA experiment~\cite{bib:fast}, and the CRAFFT telescope~\cite{bib:crafft}.
Powered by solar panels and with wireless connection, they will be deployed in a triangular arrangement with a 20\,km spacing. 

Figure~\ref{bib:fast} shows a schematic view of FAST, and the expected signal from a UHECR shower with coincidence detections at three adjacent stations.
An example of the ground coverage using 60 stations is also indicated with a comparison to Auger and TA coverages. 
To achieve an order of magnitude larger effective aperture than Auger and TA$\times$4, 500 stations are required after accounting for the standard FD duty-cycle.
The operation of a full-size FAST array will provide conclusive results on the origin and acceleration mechanism toward 100\,EeV, and open the window to charged-particle astronomy using UHECRs.

Two FAST telescopes have already been installed at the TA site, in October 2016 and September 2017~\cite{bib:fast_icrc2017}.
Both telescopes are fully remotely-operable. Both a distant vertical ultraviolet laser beam and UHECRs have been detected in time-coincidence with the Telescope Array fluorescence detector 
to confirm the sensitivity of the telescopes to UHECRs, and to demonstrate their data-taking stability during a night.

\section{Summary}
In the last decade, our understanding of UHECRs has been significantly increased by Auger and TA. 
The suppression of the energy spectrum at the highest energies is strongly confirmed. 
The mass composition using $X_{\max}$ indicates a gradual increase to a composition heavier than proton primaries above 3\,EeV.
A significant dipole structure is observed by Auger above 8\,EeV, supporting the idea of an extragalactic origin for these very high energy particles. 
Both a hotspot and a warmspot are observed close to the supergalactic plane above 57\,EeV, however there are no conclusive results yet. 
For the next decade, the upgrades of Auger and TA, AugerPrime and TA$\times$4, will provide us with a 3000\,km$^{2}$ effective area in both hemispheres, and precision measurements of UHECRs.
A future ground array requires a 30,000\,km$^2$ effective area with a sensitivity to mass composition in order to clarify the origin and acceleration mechanism toward 100\,EeV, and to establish charged-particle astronomy.
FAST is a possible solution to achieve these objectives.

\section*{Acknowledgements}
The author is very grateful to Max Malacari and Carola Dobrigkeit for a careful reading of the manuscript.
This work was supported by JSPS KAKENHI Grant Number 18H01225, 15H05443,
and Grant-in-Aid for JSPS Research Fellow 16J04564 and JSPS Fellowships H25-339, H28-4564.
This work was partially carried out by the joint research program of
the Institute for Cosmic Ray Research (ICRR) at the University of Tokyo.
\vspace{0.1cm}

\bibliography{isvhecri2018_fga}

\end{document}